# PCA-Based Relevance Feedback in Document Image Retrieval

Reza Tavoli[1], Fariborz Mahmoudi[2]

[1]Faculty of Department of Mathematics, Islamic Azad University, Chalous Branch (IAUC)
17 Shahrivar Ave., P.O. Box 46615-397, Chalous, Iran

[2]Faculty Of Department of Computer and electrical Qazvin Islamic Azad University
Qazvin, Iran

**Abstract**
Research has been devoted in the past few years to relevance feedback as an effective solution to improve performance of information retrieval systems. Relevance feedback refers to an interactive process that helps to improve the retrieval performance. In this paper we propose the use of relevance feedback to improve document image retrieval System (DIRS) performance. This paper compares a variety of strategies for positive and negative feedback**.** In addition, feature subspace is extracted and updated during the feedback process using a Principal Component Analysis (PCA) technique and based on user's feedback. That is, in addition to reducing the dimensionality of feature spaces, a proper subspace for each type of features is obtained in the feedback process to further improve the retrieval accuracy**.** Experiments show that using relevance Feedback in DIR achieves better performance than common DIR.
**Keywords:** *Relevance Feedback; Document Image; Information Retrieval; Principal Component Analysis***.**

## 1. Introduction

Document Image Retrieval System (DIRS) based on keyword spotting is performing the matching directly in the image data bypassing OCR and using word-images as queries. It is usually performed based on a comparison of common features, such as width to height ratio, word area density; shape projections features, extracted from the word document images themselves. In recent years, a number of attempts have been made by researchers to retrieval document images by word image. A detailed survey on document image retrieval up to 1997 can be found in Doermann [1]. In [2] an overview on document image retrieval system is presented. Word level image matching and retrieval has been attempted for printed documents [3-12].A key requirement for developing future document image retrieval systems is to explore the synergy between humans and computers. Relevance feedback (RF) is a technique that engages the user and the retrieval system in a process of symbiosis [14].
Relevance feedback is a powerful technique used in information retrieval systems. The idea is to adapt the system to the specific user preferences making more important weights or features that reflect the actual user needs in order to achieve higher precision. Therefore we can define relevance feedback as the process by which human and computer interact in order to automatically adjust an existing query to the real user preferences. Relevance Feedback has proven very effective for improving retrieval accuracy [13-16]. Relevance feedback refers to an interactive process that helps to improve the retrieval performance: when a user submits a query, an information retrieval system would first return an initial set of result documents and then ask the user to judge whether some documents are relevant or not; after that, the system would reformulate the query based on the user's judgments, and return a set of new results. Research has been devoted in the past few years to relevance feedback as an effective solution to improve performance of information retrieval system.

Efforts have also been made to address the problem of slow response time in content-based image retrieval, the problem being caused mainly by the high dimensionality of the feature space, typically hundreds to thousands. Mahmoudi et al [23] proposed a new feature vector for non-segmentation shape-based image indexing and retrieval. Also Shabanzade et al [24, 25] has been proposed new image indexing method for increase precision and recall in image retrieval system. Ng and Sedighian [20] made direct use of eigenimages, a method from face recognition [19], to carry out the dimension reduction. Faloutsos and Lin [18], Chandrasekaren et al. [17] and Brunelli and Mich [21] used principal component analysis (PCA) to perform dimension reduction in feature spaces. Experimental results in these works show that most real image feature sets can be considerably reduced in dimension without significant degradation in retrieval quality. However, there are two problems with the use of PCA in these works. Firstly, they adopted a fixed number for the dimension size. This strategy is questionable because for images of different complexity, the intrinsic dimensions are usually different. Secondly, the subspaces are fixed once the PCA is performed the first time and do not adapt to users' subjectivity. Generally, this kind of blind dimension reduction can be dangerous, since information can be lost if the reduction is below the embedded dimension.

In this paper, we propose the use of Relevance Feedback method to improve DIRS accuracy. Moreover, by applying the Principal Component Analysis (PCA) technique, the feature subspace is extracted and updated during the feedback process, so as to reduce the dimensionality of feature spaces, reduce noise contained in the original feature representation, and hence to define a proper subspace for each type of feature as implied in the feedback. These are performed according to positive feedbacks and hence consistently with the subjective image content. In this paper, at first we present architecture of proposed system. Then we describe the each of the box in the architecture. Each box in the architecture is an operation which includes a DIR system, mark documents and update query. In proposed method we compare a variety of strategies for positive and negative feedback which include "Only Positive Feedback", "Only Negative Feedback" and "Positive and Negative Feedback". We evaluate the proposed system with precision and recall measures. In this paper we test the proposed method on the existing database. Test results show that using relevance feedback in DIR achieve better precision and recall than common DIR.

PCA is a statistical tool for data analysis [22]. It decorrelates second order moments corresponding to low frequencies, and identifies directions of principal variations in the data. We incorporate PCA into the relevance feedback framework to extract feature subspaces in order to represent the subjective class implied in the positive feedback examples. This leads to the following benefits: 1) whitening feature distributions so that distance metrics can be defined more rationally; 2) reducing possible noise contained in the original feature representation; 3) reducing dimensionality of feature spaces, and hence 4) defining a proper subspace for each type of feature, as implied in the feedback.

This paper is organized as follows. Section 2 describes the Relevance Feedback in DIRS. Section 3 describe the Principal Component Analysis concept and Relevance feedback in the PCA feature subspace. Section 4 will show the experimental results of the proposed system. Section 5 is the conclusion.

## 2. Relevance Feedback In DIRS

In this paper, we propose the use of Relevance Feedback method to improve DIRS accuracy. System architecture is shown in Figure 1.

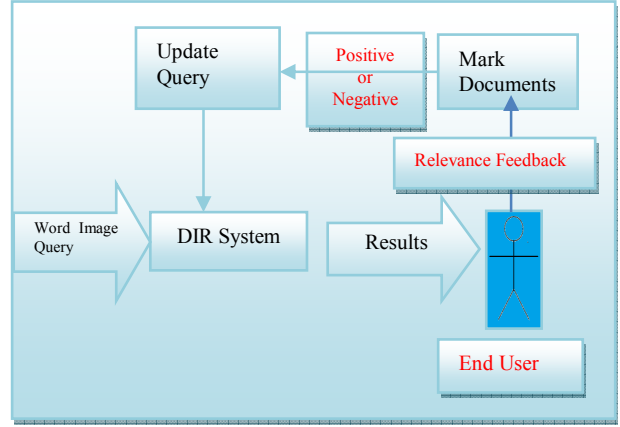

Figure 1. Proposed System

In the proposed method, at the first user enters a query. Then, the query feature vector is created. For each word block, a total of 7 different features in use, namely, Width to height ratio, Word area density, Center of gravity, Vertical projection, Top–bottom shape projections, Upper grid features and down grid features. Figure 3 depict the examples of feature vectors.

Figure 2. Examples of Feature vectors

After that, the query feature vector is compared with indexed words in the database. Minkowski distance between query feature vector and indexed words in the database is calculated [3].

$$MD(i) = \sum_{k=1}^{93} |Q(k) - W(k,i)| \quad (1)$$

$$R_i = 100 \left(1 - \frac{MD(i)}{\max(MD)}\right) \quad (2)$$

Where MD(i) is the Minkowski distance of the i word, Q(k) is the query descriptor and W(k, i) is the descriptor of the i word. Then the similarity rate of the remaining words is computed. The rate is a normalized value between 0 and 100, which depicts how similar are the words of the database with the query word. $R_i$ is the rate value of the word i, MD(i) is the Minkowski distance of the i word and max(MD) the maximum Minkowski distance found in the document database.

Then, System present retrieval results by sorting the distance according to distance measurement. Then, the user selects a set of positive and/or negative examples from the retrieved document images, and the system subsequently refines the query and retrieves a new list of documents. This paper compares a variety of strategies for positive and negative feedback which include "Only Positive Feedback", "Only Negative Feedback" and "Positive and Negative Feedback".

$$\vec{q}_m = \alpha \vec{q}_0 + \beta \frac{1}{|W_r|} \sum_{\vec{w}_j \in W_r} \vec{W}_j \qquad (3)$$

For Negative feedback, we select as non-relevant all the word images from the initial query result which the user judged to be non-relevant. For negative feedback Rocchio's formula is changed to:

$$\vec{q}_m = \alpha \vec{q}_0 - \gamma \frac{1}{|W_{nr}|} \sum_{\vec{w}_j \in W_{nr}} \vec{W}_j \qquad (4)$$

For Positive and Negative feedback, we select as relevant and non-relevant all the word images from the initial query result which the user judged to be relevant and non-relevant. For Positive and negative feedback Rocchio's formula is changed to:

$$\vec{q}_m = \alpha \vec{q}_0 + \beta \frac{1}{|W_r|} \sum_{\vec{w}_j \in W_r} \vec{W}_j - \gamma \frac{1}{|W_{nr}|} \sum_{\vec{w}_j \in W_{nr}} \vec{W}_j \qquad (5)$$

In Equation (4),(5) and (6), $q_0$ is the original query vector, $W_r$ and $W_{nr}$ are the set of known relevant and non-relevant words in documents respectively, and α, β, and γ are weights attached to each term.

## 3. Relevance Feedback in the PCA feature subspace

The other major contribution in the proposed relevance feedback approach to document image retrieval is to apply the principal component analysis technique to select and updated a proper feature subspace during the feedback process. This algorithm extracts more effective, lower-dimensional features from the originally given ones, by constructing proper feature subspaces from the original spaces, to improve the retrieval performance in terms of speed, storage requirement and accuracy. In this section, we first present the PCA algorithm, followed by a detailed description of how we apply PCA in relevance feedback in content-based image retrieval.

### 3.1 Principal Component Analysis (PCA)

Consider an ensemble of n-dimensional vectors $\{x = [x_1,...,x_n]^T\}$ whose distribution is centered at the origin $E(x) = 0$. The covariance between each pair of variable is $r_{ij} = E\{(x_i - \bar{x}_i)(x_j - \bar{x}_j)\} = E\{x_i x_j\}$, where E is the expectation operator. The parameters $r_{ij}$ can be arranged to form the $n \times n$ covariance matrix

$$R_x = E\{(x - \bar{x})(x - \bar{x})^T\} = E\{xx^T\}. \qquad (6)$$

Assuming $\det(R_x) \neq 0$, then by applying eigenvector decomposition, $R_x$ can be decomposed into the product of three matrices

$$R_x = W \Lambda W^{-1} \qquad (7)$$

Where $\Lambda = diag\{\lambda_1,...,\lambda_n\}$ are the eigenvalues and $W = [w_1,...,w_n]^T$ are the corresponding eigenvectors. $W$ is orthogonal in that $WW^T = I$. So the columns of $W$ form a new orthogonal basis that is a linear transformation from of the original basis.

The eigenvector decomposition can be used to whiten the feature distributions as follows. Project the original feature vectors $x$ onto the eigenvector basis (without dimension reduction), obtaining the coordinates $x$, which is equivalent to rotating the feature basis; then rescale the coordinates by the factor of $1/\sqrt{\lambda_j}$ to obtain the whitened feature vector $y$ and $y = Wx$. After the whitening, we are able to calculate the Mahalanobis distance between $x_1$ and $x_2$ in the original feature space by the simple Euclidean distance between the corresponding $y_1$ and $y_2$, in the whitened feature space, i.e.

$$dist(x_1, x_2) = \frac{(x_1 - x_2)^T}{\sum(x_1 - x_2)} = \|y_1 - y_2\|. \qquad (8)$$

If we only select the first eigenvectors as the orthonormal basis vectors to form a subspace $L = Span(W')$, then any vector $x$ in the original space can be linearly transformed to with the new representation $y'$

$$y' = W'x. \qquad (9)$$

An approximation to the original can be reconstructed from the projection $y'$ as $x' = W'^T_{y'} = W'^T W'x$. The mean squared reconstruction error is

$$J_e(M) = E\{\|x - x'\|^2\} = \sum_{i=m+1}^{n} \lambda_i. \qquad (10)$$

We can choose the set of eigenvectors used for the reconstruction to minimize this: Sort eigenvalues in descending order so that $\lambda_i > \lambda_j > 0$, where $i>j$; this also sorts the corresponding eigenvectors in the descending order of their significance. The mean square reconstruction error $J_e$ can thus be minimized [22].

We can choose the set of eigenvectors used for the reconstruction to minimize this: Sort eigenvalues in descending order so that $\lambda_i > \lambda_j > 0$, where $i>j$; this also sorts the corresponding eigenvectors in the descending order of their significance. The mean square reconstruction error $J_e$ can thus be minimized [22].

### 3.2 Relevance Feedback in the PCA Feature Subspaces

As described in the last subsection, PCA can be used to reduce the dimensionality of the feature space, and to extract the principal lower-dimensional subspace of the original feature space. A reasonable deduction in dimensionality causes little decrease in performance.

The following describes the PCA embedded algorithm.

1) Initialization of System: For each feature type, do the following:
   a. Perform PCA on all the word images in the original feature space, obtaining the eigenvalues and the corresponding eigenvectors calculated by (6) and (7). Sort the eigenvectors in the order of descending eigenvalues.
   b. Select Subset of eigenvectors as a basis vectors.
   c. Convert Data from Original feature space to new feature Space according to eigenvalues.
2) Retrieval and Feedback
   a. Enter Query
   b. Convert the query feature vector to new query feature vector according to eigenvalues.
   c. Calculate the distance between all word images in the database and the current query in the dimensional feature subspace.
   d. Sort by distances and provide the new ranking list to the user.

## 4. Experimenting the proposed system

In our experiments, the evaluation of the proposed system was based on 100 document images. The database of the documents has been created automatically from various digital text documents. In order to calculate the precision and recall values 30 searches were made using random words. In this paper, we tested results with variety of strategies for positive and negative feedback. In positive feedback strategy, we set Rocchio's formula with α=1 and β=0.82. In positive feedback, the precision values obtained are depicted in Figure 4.

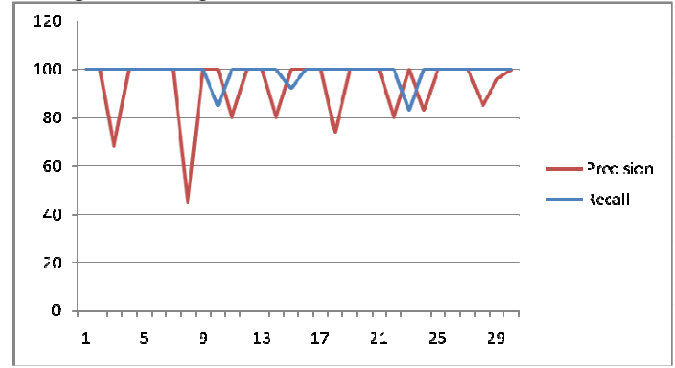

Figure 3. The variation of the precision and Recall Coefficient of the proposed method(Positive Feedback) for 30 searches. The Average Precision is 93.03% and Average Recall is 98.66%

As shown in Figure 3, by using positive feedback in DIRS, performance of DIRS in term of average precision is increased and term of average of recall is fixed. Positive feedback turns out to be much more valuable than negative feedback, and so most IR systems set γ<β. In this system, we have a few judged documents, and then we would like a higher α and β. Experiments show that using positive Feedback in DIR achieves better performance than common DIR.

Table 1 depicts comparison the average precision and average recall of the approach with DIRS [3] and WDIRS [4].

TABLE I. COMPARISON THE AVERAGE PRECISION AND RECALL BETWEEN PROPOSED SYSTEM AND DIRS AND WDIRS

| Methods | Precision | Recall |
|---|---|---|
| DIRS[3] | 87.8% | 99.26% |
| WDIRS[4] | 55.43% | 94.78% |
| Positive Feedback In DIRS | 93.03% | 98.66% |

As shown in Table 1, the average precision in WDIRS and DIRS is 55.43% and 87.8%, respectively. Also, average recall in WDIRS and DIRS is 94.78% and 99.26%, respectively. After using positive feedback method in DIRS the average precision is 93.03% and average recall become 98.66% respectively.

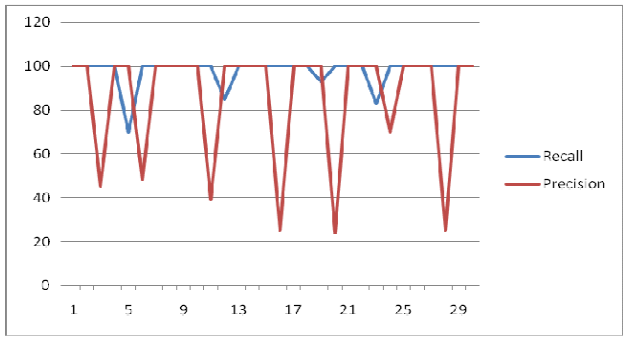

Figure 4. The variation of the precision and Recall Coefficient of the proposed method(Negative Feedback) for 30 searches. The Average Precision is 85.86% and average Recall is 97.7%

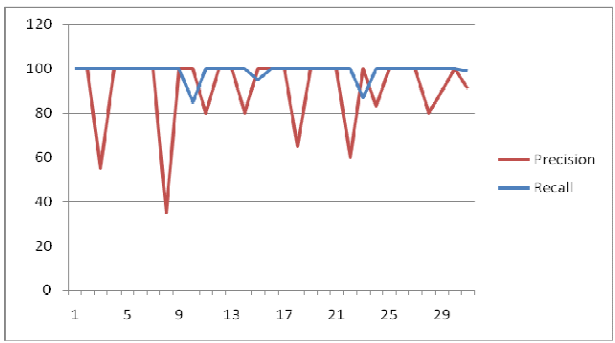

Figure 5. The variation of the precision and Recall Coefficient of the proposed method(Positive and Negative Feedback) for 30 searches. The Average Precision is 90.93% and average Recall is 98.9%

As shown in Figure 6, by using negative feedback in DIRS, performance of DIRS in term of average precision and recall is decreased. Because Negative relevance feedback is a special case of relevance feedback where we do not have any positive example; this often happens when the search results are poor.

In figure 6, by using positive and negative feedback in DIRS, performance of DIRS in term of average precision is increased and term of average of recall is fixed.

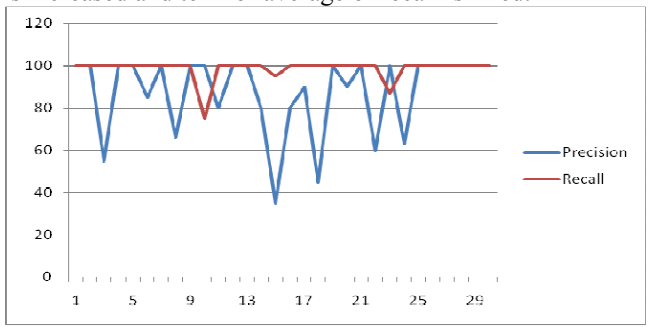

Figure 6. The variation of the precision and Recall Coefficient of the proposed method(in the PCA Feature Subspace) for 30 searches. The Average Precision is 87.6% and average Recall is 98.5%

Figure 8 depict the using perform PCA in retrieval process, performance of DIRS in term of average precision and recall is decreased. A reasonable deduction in dimensionality causes little decrease in performance. This is especially true in content-based image retrieval since the components removed from the original image feature space often correspond to noise. According to our experimental results on a large amount of data, dropping 80% of the feature dimensions leads to only about 5% reconstruction error; dropping 90% dimensions gives only about 10% reconstruction error. Yet the retrieval speed has been improved significantly as a result of such dimension reductions. There are two advantages of using PCA: 1) dimension reduction is achieved; 2) noise reduction is achieved.

## 5. Conclusions

In many information retrieval systems relevance feedback is used to increase accuracy. In this paper we use the relevance feedback technique to improve document image retrieval system performance. This paper compares a variety of strategies for positive and negative feedback. These are "Only Positive Feedback", "Only Negative Feedback" and "Positive and Negative Feedback". Experiment results show that using relevance Feedback in DIR achieves better performance than common DIR.

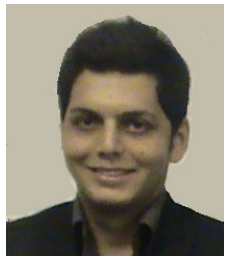

**Reza Tavoli** received his B.s in software engineering from Iran University of Science & Technology, Behshahr, Iran. He received his M.s software engineering from Islamic Azad University, science & Research Branch, Tehran, Iran. Currently, He is pursuing PhD in Software engineering at Islamic Azad University, Qazvin Branch, and Qazvin, Iran. His research interests Include document image retrieval and data mining.

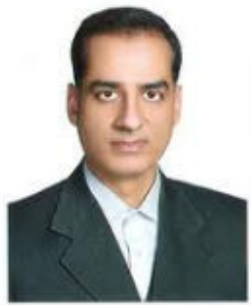

**Fariborz Mahmoudi** is an Assistant Professor at Qazvin Islamic Azad University, Qazvin, Iran. He received his B.s in software engineering From Shahid Beheshti University, Tehran, Iran. He received his M.s From in Computer engineering from Amir Kabir University, Tehran, Iran. His research interests Include image retrieval.